\documentclass[aps,prb,twocolumn,superscriptaddress]{revtex4}
\usepackage{amsmath,amssymb}
\usepackage{graphicx}
\usepackage{epstopdf}
\usepackage{bm}
\usepackage{color}
\usepackage{mhchem}

\newcommand{\li}{$^{7}$Li}
\newcommand{\pho}{$^{31}$P}

\newcommand{\tno}{$T_N^\text{olivine}$}

\newcommand{\tetra}{LiCoPO$_4^\text{tetra}$}
\newcommand{\olivine}{LiCoPO$_4^\text{olivine}$}
\newcommand{\lcp}{LiCoPO$_4$}

\newcommand{\slr}{$T_1^{-1}$}
\newcommand{\slrp}{$^{31}T_1^{-1}$}
\newcommand{\slrli}{$^{7}T_1^{-1}$}

\begin{document}

\title{Unusual spin fluctuations and magnetic frustration in 
olivine and non-olivine LiCoPO$_4$ detected by $^{31}$P and  
$^7$Li nuclear magnetic resonance}

\author{S.-H. Baek}
\email[]{sbaek.fu@gmail.com}
\affiliation{IFW-Dresden, Institute for Solid State Research,
PF 270116, 01171 Dresden, Germany}
\author{R. Klingeler}
\affiliation{Kirchhoff Institute for Physics, University of Heidelberg, 
69120 Heidelberg, Germany} 
\author{C. Neef}
\affiliation{Kirchhoff Institute for Physics, University of Heidelberg, 
69120 Heidelberg, Germany} 
\author{C. Koo}
\affiliation{Kirchhoff Institute for Physics, University of Heidelberg, 
69120 Heidelberg, Germany} 
\author{B. B\"{u}chner}
\affiliation{IFW-Dresden, Institute for Solid State Research,
PF 270116, 01171 Dresden, Germany}
\affiliation{Institut f\"ur Festk\"orperphysik, Technische Universit\"at 
Dresden, 01062 Dresden, Germany}
\author{H.-J. Grafe}
\affiliation{IFW-Dresden, Institute for Solid State Research,
PF 270116, 01171 Dresden, Germany}
\date{\today}

\begin{abstract}
We report \pho\ and \li\ nuclear magnetic resonance (NMR) studies in new 
non-olivine LiZnPO$_4$-type \tetra\ microcrystals, where the Co$^{2+}$ ions 
are tetrahedrally  
coordinated.  Olivine LiCoPO$_4$, which was directly transformed from \tetra\   
by an annealing process, was also studied and compared.  The uniform bulk 
magnetic susceptibility  
and the \pho\ Knight shift obey the Curie-Weiss law for both materials with 
a high spin Co$^{2+}$ ($3d^7$, $S=3/2$), but the Weiss 
temperature $\Theta$ and the effective magnetic moment $\mu_\text{eff}$ are 
considerably smaller in \tetra.  The spin-lattice 
relaxation rate \slr\ reveals a quite different nature of 
the spin dynamics in the paramagnetic state of both materials. Our NMR results 
imply that strong  
geometrical spin frustration  
occurs in tetrahedrally coordinated LiCoPO$_4$, which may lead to the 
incommensurate magnetic ordering.

\end{abstract}


\maketitle

\section{Introduction}

The olivine structured lithium transition metal phosphates Li$M$PO$_4$ ($M$ = Fe, Mn, 
Co, and Ni) have attracted interest from both fundamental and technical points 
of view. They reveal a variety of unusual magnetic, magnetoelectric, and 
ferrotoroidic properties associated with high spin (HS) $M^{2+}$ 
ions.\cite{vaknin04,vanaken07,aken08,jensen09a,toft-petersen12} Also,  
their low cost, low toxicity, high stability, and high energy density made them promising 
candidates of high-voltage cathode materials for Li-ion 
batteries.\cite{padhi97,chung02,fisher08,bramnik08,kang09,murugan09}
Among the olivine Li$M$PO$_4$ family, LiCoPO$_4$ features a very large 
linear magnetoelectric effect\cite{rivera94,kornev00}
and a high theoretical energy density up to 801 Wh/kg based on its  
high discharge plateau 4.8 V versus Li/Li$^+$.\cite{amine00,okada01} 

\begin{figure}
\centering
\includegraphics[width=0.6\linewidth]{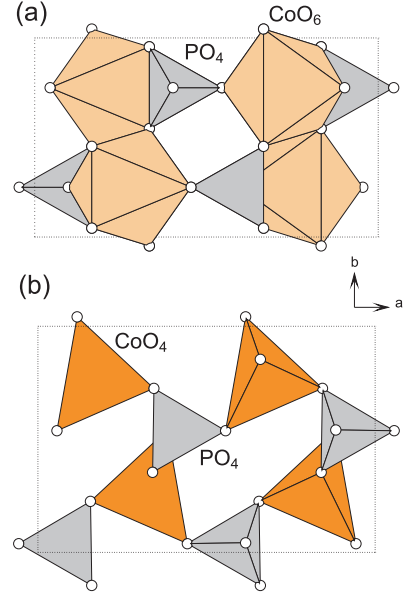}
\caption{\label{fig:structure} (Color online)
Crystal structure of (a) olivine ($Pnma$ symmetry) and (b) tetrahedral ($Pn2_1a$ 
symmetry) LiCoPO$_4$, projected along the crystallographic $c$ axis. Li atoms 
are omitted for clarity.   Their 
unit cells are drawn as dotted lines.   
}  
\end{figure}

Olivine LiCoPO$_4$ crystallizes in the orthorhombic $Pnma$ space 
group.\cite{santoro66}  The Co$^{2+}$  
($S=3/2$) ions sit in the center of the distorted CoO$_6$ octahedra which share corners and 
edges with PO$_4$ tetrahedra, as illustrated in Fig. \ref{fig:structure}(a).
Below $T_N\sim 21$ K, the moments order antiferromagnetically, with a tilt of  
4.6$^\circ$ away from the crystallographic $b$ axis within the $bc$ plane,  
and the crystal structure changes to the $P12'_11$ symmetry.\cite{kharchenko01,vaknin02,  
szewczyk11} In this low  
symmetry, a nonzero toroidal moment is allowed and confirmed 
experimentally.  \cite{vanaken07,ederer07}

For optimal performance of electrode materials for Li-ion batteries,
it is crucial to get insight into the fundamental structural, electronic, and 
magnetic properties of the materials and their limits and trends for 
applications.\cite{chernova11}
Therefore, recently discovered non-olivine structured LiCoPO$_4$ is 
interesting and may provide an important step forward in the understanding of 
the impact of structure and magnetism on the performance of a battery 
material.  \tetra\ possesses $Pn2_1a$  
symmetry and consists of CoO$_4$ tetrahedra, instead of CoO$_6$ octahedra in the 
olivine structure, sharing only corners with PO$_4$ 
tetrahedra\cite{hautier11,jahne13} [see Fig. \ref{fig:structure}(b)].  
The non-olivine structure becomes unstable at high temperatures towards the olivine one.  

Initial studies indicate that tetrahedral LiCoPO$_4$ exhibits poor  
performance in terms of cycling stability and discharge capacity compared to 
the olivine phosphate.\cite{jahne13}
In an attempt to elucidate the detailed magnetic properties of both compounds 
associated with their structural aspects, 
we carried out an NMR study on 
tetrahedral \tetra\ as well as on the annealed olivine compound.   
Our results of the Knight shift and the spin-lattice relaxation rates show 
that the tetrahedrally coordinated Co$^{2+}$ spins in \tetra\  
are strongly frustrated, resulting in quite different magnetic properties,
compared to olivine LiCoPO$_4$. The magnetic structure in the ordered state is 
likely incommensurate.

\section{Sample preparation and experimental details}

Non-olivine \tetra\ microcrystals with $Pn2_1a$ symmetry were synthesized by the
microwave-assisted hydrothermal synthesis technique, as described in detail in 
Refs. \onlinecite{jahne13,neef13}. Olivine LiCoPO$_4$ with $Pnma$ symmetry was 
obtained by annealing the non-olivine compound at 700$^\circ$C for 24 hours 
under an argon atmosphere.   

The temperature dependence of the static uniform magnetic susceptibility $\chi(0,0)$ 
of \tetra\ and \olivine\ was measured using a superconducting quantum 
interference device (SQUID) magnetometer in the field 
of 1 kOe after cooling in zero magnetic field.

\li\ and \pho\ NMR measurements have been carried out using a spin-echo method 
in a fixed field of 7.0494 T in the temperature range of 5--400 K. 
Since \pho\ (nuclear spin $I=1/2$, $\gamma_n=17.2356$ MHz/T) does not involve 
electric quadrupole  
effects, it is an ideal probe to study magnetism and spin 
fluctuations in these materials.  
\pho\ NMR spectra over most of the measured temperature range are relatively narrow 
and quite symmetric, in comparison with the large linewidth and strong 
anisotropy observed in other 
Li phosphates,\cite{arcon04,rudisch13} which allowed us  
to determine the Knight shift and the linewidth reliably.
The spin-lattice relaxation rates \slr\ of both $^{31}$P and $^7$Li 
($I=3/2$, $\gamma_n=16.5471$ MHz/T) were measured using the  
saturation recovery method and $T_1$ 
was obtained by fitting the relaxation of the nuclear magnetization $M(t)$ to 
a single exponential function, $1-M(t)/M(\infty)=A\exp(-t/T_1)$ where $A$ is a 
fitting parameter.

\section{Experimental results and discussion}

\subsection{Magnetic susceptibility $\chi$ and \pho\ Knight shift $\mathcal{K}$}
 
\begin{figure}
\centering
\includegraphics[width=\linewidth]{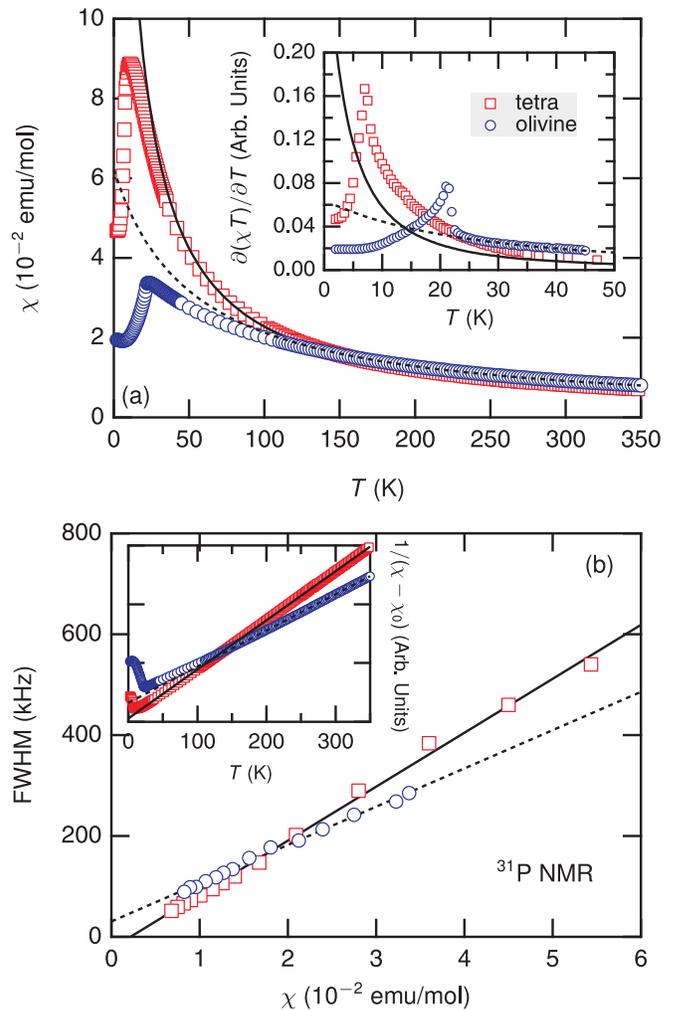}
\caption{\label{fig:chi} (Color online)(a) Temperature dependence of static susceptibility 
and magnetic specific heat (inset) of both olivine 
and tetrahedral LiCoPO$_4$. Solid curves are Curie-Weiss fits. (b) NMR 
linewidth (FWHM) of $^{31}$P spectrum (see Fig. \ref{fig:31spec}) tracks 
$\chi$ in the whole temperature range investigated. Inset: Inverse static susceptibility.}
\end{figure}

Figure \ref{fig:chi} shows the molar static susceptibility $\chi_\text{mol}=M/H$ 
as a function of temperature for both olivine and tetrahedral LiCoPO$_4$. The 
drop of the $\chi$ data at low temperatures indicates that in both systems 
long range antiferromagnetic order evolves at low temperature. The transition 
can be clearly observed in the magnetic 
specific heat $c_\text{mag} \sim \partial(\chi_\text{mol} T)/\partial T$ (see 
inset) as both compounds show an anomaly at 
$T_N$ = 21 and 7 K, respectively.

At high temperatures, the data in both compounds follow a Curie-Weiss (CW) law, 
$\chi_\text{mol}=C/(T+\Theta)+\chi_0$, where 
$\chi_0$ is a $T$-independent susceptibility [see inset of 
Fig.~\ref{fig:chi}(b)]. Fitting the data with the CW law 
yields the Weiss temperatures $\Theta=52(5)$ K (olivine) and $\Theta=7(1)$ K 
(tetra), and the effective magnetic 
moments $\mu_\text{eff} = 5.1 \mu_B$ (olivine) and $4.4 \mu_B$ (tetra). Both 
values of $\mu_\text{eff}$ exceed the 
spin-only value of 3.87 $\mu_B$ expected for the HS 3$d^7$ configuration for 
full quenching of the orbital moment. In \tetra, the 
orbital admixture to the measured effective $g = 2.27$ is governed by the 
reduced spin-orbit coupling $\lambda$ and the 
tetrahedral crystal field (CF) splitting $\Delta_t$ of the Co$^{2+}$ orbital 
states of $e_g$ and  
$t_{2g}$ symmetry. It can be approximated by 
 
\begin{equation}
g = 2-\frac{8\cdot \lambda}{\Delta_t}.
\end{equation}
 
With $\lambda \approx -143$ cm$^{-1}$,\cite{sati06} the data are consistent 
with $\Delta_t \approx 4250$ cm$^{-1}$. In contrast, the 
electronic configuration of HS Co$^{2+}$ in the octahedral configuration 
$t_{2g}^5e_g^2$ realized in \olivine\
exhibits a stronger orbital contribution since it involves only partially filled 
low-lying orbital triplet states with 
pseudo angular momentum $\tilde{l}=1$. The observed effective moment agrees 
well with the findings in Ref.~\onlinecite{vaknin02}.
 
\begin{figure}
\centering
\includegraphics[width=\linewidth]{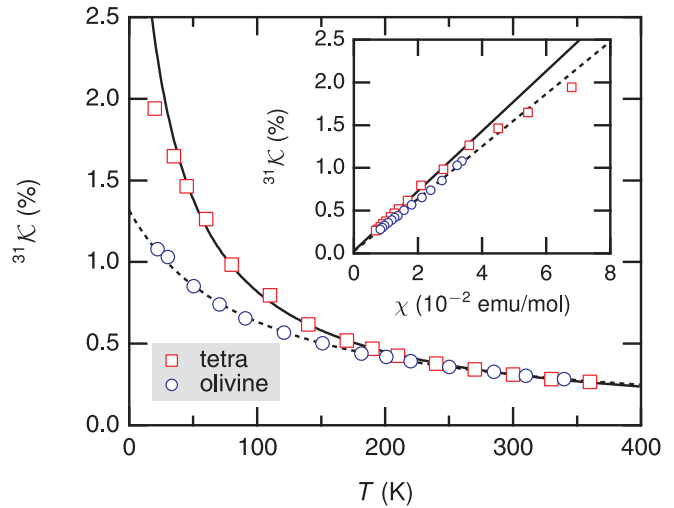}
\caption{\label{fig:31K} (Color online) The $^{31}$P Knight shift  as a function of 
temperature. Solid curves are Curie-Weiss fits. 
The inset shows a plot of $\mathcal{K}(T)$ versus $\chi(0,0)(T)$ with $T$ as 
implicit parameter.  The slope of the data 
yields the hyperfine coupling constants $A_\text{hf}=1.96$ kOe/$\mu_B$ and 
1.71 kOe/$\mu_B$ for \tetra\ and \olivine, 
respectively. The $T$-independent orbital shifts are found to be  nearly zero (see text). }
\end{figure}

In this paper, we mainly concentrate on the NMR data obtained on both 
materials. Hence, we note that the broadening of the NMR line is essentially 
determined by the uniform bulk susceptibility as can be deduced from Fig. 
\ref{fig:chi} (b) where the full width at half maximum (FWHM) of the \pho\ NMR 
spectra is plotted against $\chi$ with $T$ as an implicit 
parameter. Further information on the static magnetic properties is obtained 
from measurements of the \pho\ Knight shift, $^{31}\mathcal{K}$. Its 
temperature dependence confirms CW-behavior in both materials. The data in
Fig.~\ref{fig:31K} are well described by 
$\mathcal{K}=C'/(T+\Theta)+\mathcal{K}_0$, where $\mathcal{K}_0$ is the 
$T$-independent orbital shift. For both materials, $\mathcal{K}_0$ appears 
negligibly small, indicating that 
$\mathcal{K}$ probes almost entirely the spin part of the magnetic susceptibility.
 
 
Focusing on the evolution of antiferromagnetic order, the static 
susceptibility data imply deviations from the 
mean-field Curie-Weiss-like behavior already at relatively high temperatures. 
In the olivine sample, small deviations from 
the experimental data are observed below $\sim 250$\,K as visible in 
Figs.~\ref{fig:chi}(a) and (b, inset). The 
experimentally observed susceptibility is smaller than predicted by the 
Curie-Weiss law which is in agreement with the 
evolution of antiferromagnetic fluctuations well above $T_N$. In the tetragonal 
polymorph, such behavior is present 
below $\sim 50$\,K which is best seen if the magnetic specific heat in Fig. 
\ref{fig:chi}~(a) (inset) is considered.

 

\begin{figure}
\centering
\includegraphics[width=\linewidth]{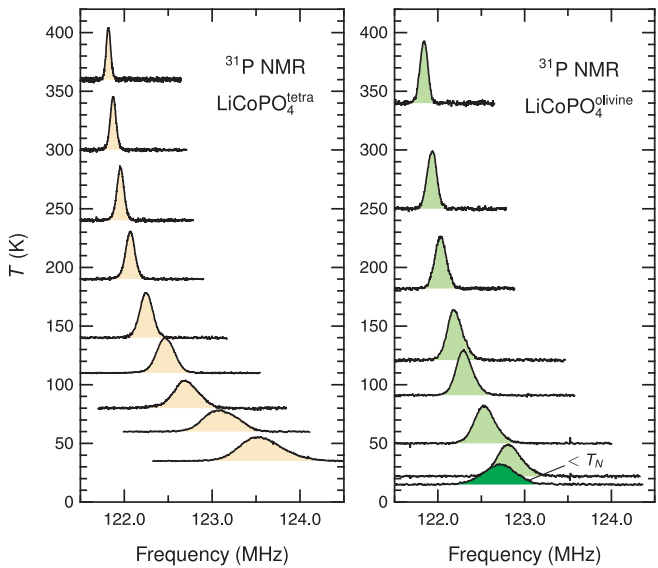}
\caption{\label{fig:31spec} (Color online) Temperature evolution of $^{31}$P NMR spectrum 
obtained at 7.0494 T for both tetrahedral (left) and olivine (right) LiCoPO$_4$ 
in the paramagnetic state. The spectrum in the ordered state for 
\olivine\ is also shown for comparison.}  
\end{figure}

In order to compare the static susceptibility measured by bulk and local 
techniques, the inset of Fig. \ref{fig:31K} 
shows a plot of $^{31}\mathcal{K}$ versus $\chi$.  $^{31}\mathcal{K}$ is 
proportional to $\chi$ over a wide temperature 
range. The small deviation from linearity occurring below $\sim60$ K for \tetra\ 
is attributed to a small amount of 
paramagnetic (PM) impurities to which $\mathcal{K}$ is insensitive. This 
account is indeed corroborated by the good 
agreement between the linewidth and $\chi$ in the $T$ region where 
$\mathcal{K}$ deviates $\chi$, as shown in the inset 
of Fig. \ref{fig:chi}, since a random distribution of PM moments would broaden 
the NMR line without affecting its shift.
 
The linear slope of $d\mathcal{K}/d\chi$ corresponds to the hyperfine (hf) 
coupling constants, $A_\text{hf}=1.96$ 
kOe/$\mu_B$ (tetra) and 1.71 kOe/$\mu_B$ (olivine). Note that 
$^{31}\mathcal{K}$ in both compounds almost vanishes at $\chi =\chi_0\sim0$, confirming the 
non-spin susceptibility contribution $\mathcal{K}_0\sim0$.   


\subsection{Spin-lattice relaxation rate \slr\ and dynamical susceptibility}

\begin{figure}
\centering
\includegraphics[width=\linewidth]{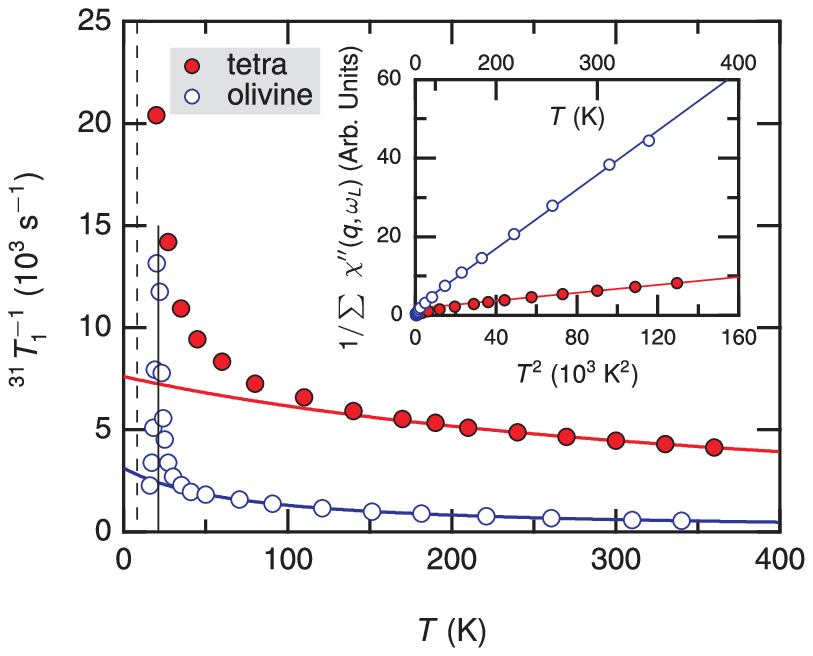}
\caption{\label{fig:31invT1} (Color online) $^{31}$P spin-lattice relaxation rate \slrp\ as a 
function of temperature in olivine and tetrahedral LiCoPO$_4$ samples. For 
\olivine, \slrp\ reveals a sharp 
magnetic transition at $T_N=21$ K (vertical solid line), while  it 
diverges towards $T_N=7$  
K (vertical dashed line) for \tetra. Solid curves are equivalent to the linear 
fit in Fig. \ref{fig:T1comp}.   
The inset presents a plot of $1/\sum_q\chi''(q,\omega_L)$ versus $T^2$, 
to show the inverse dynamical 
susceptibility that is quadratic in temperature.} 
\end{figure}

The $^{31}$P  spin-lattice relaxation rate  
\slrp\ as a function of temperature is presented in Fig. \ref{fig:31invT1}.  
For the olivine compound, \slrp\ increases steadily with decreasing $T$ and is 
rapidly enhanced below 40 K, exhibiting a sharp anomaly followed by a rapid 
drop. The sharp peak (vertical solid line) indicates the onset of magnetic 
order at $T_N=21$ K, which agrees with literature values.\cite{santoro66,vaknin02,szewczyk11}
The rapid upturn above $T_N$, which is
due to the critical slowing down of spin fluctuations towards 
magnetic order, is a rough measure of dimensionality of the magnetic system as well.  
In our case, the relatively sharp  
transition width which is comparable to $T_N$ indicates the 
quasi-three-dimensional nature of  
the magnetic order rather than two-dimensional (2D).\cite{vaknin02,baker11}
For \tetra, the \slrp\ data are an order of magnitude larger than those of 
\olivine, displaying a similar temperature dependence.  
$T_N$ could not be identified since $^{31}T_1$
becomes too short to be measured near the transition, 
but the temperature at which \slrp\ diverges appears to be consistent with $T_N=7$ 
K (vertical dashed line).

For olivine \lcp, inelastic neutron scattering yields moderate magnetic 
exchange coupling constants.\cite{tian08} 
In the $bc$-plane, the dominating nearest neighbor coupling amounts to 
$J_{\|}^{nn} = 9$\,K while smaller next nearest 
neighbor and interlayer couplings of $\sim 1 - 2$\,K imply a tendency to weak 
frustration and 2D behavior. In the case 
of \tetra, where $T_N$ as well as $\Theta$ are significantly smaller than in 
the olivine material, magnetic coupling is 
presumingly weaker and/or the tendency towards 2D and frustration stronger.

In the paramagnetic  
limit, \slr\ could be approximated by the relation,\cite{moriya56a}
\begin{equation}
\label{eq:2}
T_1^{-1}\propto \frac{A_\text{hf}^2\sqrt{S(S+1)}}{\hbar J_\text{ex}}.
\end{equation}
Using this equation, one can estimate the ratio, 
$T_1^{-1}\text{(tetra)}/T_1^{-1}\text{(olivine)}=6.5$, using the relation 
$J_\text{ex}\equiv 3k_B\Theta/[z S(S+1)]$ where $z$ is the number of 
nearest-neighbors. This value satisfactorily
accounts for the difference of \slrp\ between the two compounds at high 
temperatures, proving that the system indeed lies in the localized limit at 
high temperatures.  

\begin{figure}
\centering
\includegraphics[width=\linewidth]{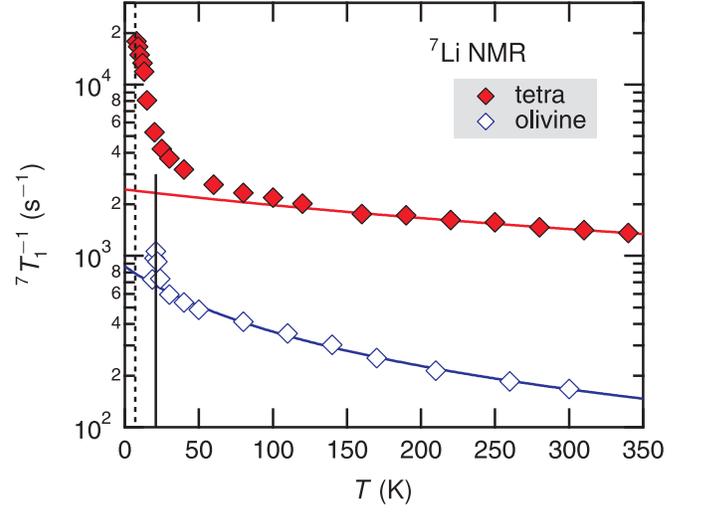}
\caption{\label{fig:7invT1} (Color online) Temperature dependence of the $^{7}$Li 
spin-lattice relaxation rate.  
Data in the PM region resemble those of \slrp, indicating that both \li\ and 
\pho\ nuclei are governed by the same relaxation mechanism. The solid curves 
are identical with those in Fig. \ref{fig:31invT1} with scaling.  
A sharp anomaly was detected at $T_N=21$ K for \olivine, similar to the case of \pho. }
\end{figure}

In general, while the Knight shift is proportional to the static spin 
susceptibility at $q=0$, i.e. 
$\mathcal{K}=A_\text{hf}\chi(0,0)$,  \slr\ reflects the $q$-average of the imaginary 
part of the dynamical  
susceptibility $\chi''$ at low energy,\cite{moriya63} 
\begin{equation}
\label{eq:T1}
T_1^{-1} \propto T \gamma_n^2 A_\text{hf}^2 \sum_q \chi''(q,\omega_L)/\omega_L,
\end{equation}
where $\gamma_n$ is the nuclear gyromagnetic ratio and $\omega_L$ the Larmor 
frequency.
Since there is no difference of the Knight shift at high temperatures far 
above $T_N$ between the two compounds, we conclude that spin fluctuations are 
of dominantly antiferromagnetic nature for both olivine and tetrahedral LiCoPO$_4$.  

\begin{figure}
\centering
\includegraphics[width=0.9\linewidth]{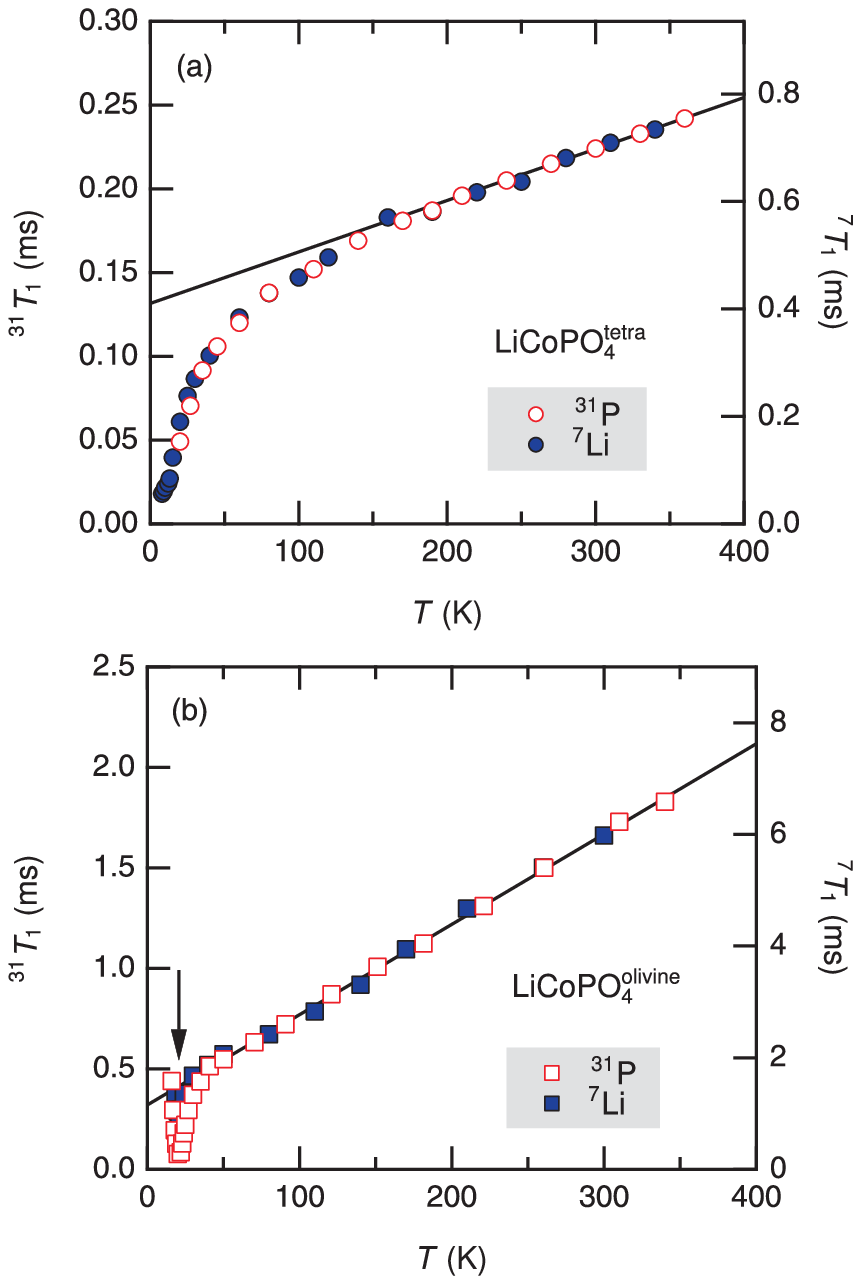}
\caption{\label{fig:T1comp} (Color online) For (a) \tetra\ and (b) \olivine, the spin-lattice 
relaxation times $^{31}T_1$ and $^7T_1$ as a function of temperature
are scaled to each other, demonstrating that 
they detect the dynamical susceptibility 
$\sum_q \chi''(q,\omega_L)$ of the materials.
While the linear $T$ dependence of $T_1$ is well maintained over the whole 
temperature range investigated, except near the transition at $T_N$ for \olivine,
it breaks down below $\sim150$ K for \tetra.  
}
\end{figure}

The most striking feature is that \slrp\ for both compounds increases with 
decreasing temperature, as shown in Fig. \ref{fig:31invT1}. Such a $1/T$ dependence 
of \slr\ is very  
rare in the paramagnetic limit.   
Figure \ref{fig:T1comp} clearly 
shows the linear variation of $T_1$ in terms of $T$, particularly, in \olivine.  
This in turn implies that the $q$-average of the
dynamical susceptibility  
$\sum_q\chi''(q,\omega_L)$ from Eq. (\ref{eq:T1}) varies in proportion to 
$1/T^2$, in contrast to the uniform static susceptibility $\chi(0,0)$ that 
obeys the CW law.   
A plot of $1/\sum_q\chi''(q,\omega_L)$ versus $T^2$ is given in the inset of Fig. 
\ref{fig:31invT1}, which provides evidence of the quadratic  
temperature dependence of the inverse dynamical susceptibility. Note that this 
plot eliminates the effect of the hf coupling constants, allowing the direct 
comparison of the two systems.  

To ensure that the unusual $T$ dependence of \slrp\ is not 
site dependent,  but represents the intrinsic dynamical susceptibility of the system, we 
also measured the  
\li\ spin-lattice relaxation rate,  
\slrli, as a function of temperature.  The results are presented in Fig. 
\ref{fig:7invT1}, revealing a $T$ dependence similar to \slrp.  
In fact, Fig. \ref{fig:T1comp} proves that $^{31}T_1$ and $^{7}T_1$ as a function of 
temperature are accurately  
scaled to each other for both compounds.   
\footnote{From the scaling, one can deduce the hf coupling constants of \li\ 
directly from those of \pho\ ; 0.79 kOe/$\mu_B$ (tetra) and 
0.69 kOe/$\mu_B$ (olivine).}

From Figs. \ref{fig:31invT1} and \ref{fig:T1comp},
the different behaviors of \slr\ in the two compounds are noticeable.  
Namely, for \olivine, the $T_1$ data follow a linear $T$ behavior in the whole 
temperature range investigated except the region near \tno. In contrast, for 
\tetra, the $T_1$ data deviate 
from the $T$-linear behavior at $\sim150$ K, suggesting that an additional 
relaxation mechanism is developed.  
Another noticeable feature is that the $T$-linearity  
of $T_1$ is considerably smaller in \tetra, seemingly approaching the normal CW behavior 
[i.e., $T_1(T) \rightarrow \text{ constant}$ at high $T$] which is observed in 
other olivine lithium phosphates, LiMnPO$_4$ and LiFePO$_4$.\cite{arcon04}  

The different magnetic properties of \tetra\ and \olivine\ could be 
understood by considering their inherent spin networks.   
There are five exchange pathways between the Co$^{2+}$ spins in olivine 
LiCoPO$_4$.\cite{dai05,tian08} 
In the $bc$ plane, one can identify the 
nearest-neighbor coupling $J_1$ mediated
through Co-O-Co superexchange path and the next-nearest-neighbor coupling 
$J_2$ and $J_3$ along the $b$ and $c$ axes, respectively, through PO$_4$ 
tetrahedra [Fig. \ref{fig:structure}(a) depicts only $J_1$ coupled Co atoms]. 
The interplane couplings $J_4$ and  
$J_5$ are also mediated by PO$_4$ tetrahedra and are known to be 
ferromagnetic, while the intraplane couplings $J_1$, $J_2$, and $J_3$ are all 
antiferromagnetic. This 
spin network involves weak geometrical frustration since  
$J_1$ is much larger than other exchange couplings.\cite{tian08}  

This situation is dramatically altered in \tetra.    
As clearly shown in Fig. \ref{fig:structure}(b), there is no longer a Co-O-Co superexchange 
path and all the exchange interactions are mediated  
by corner-shared PO$_4$ tetrahedra and might be comparable to each other in strength.   
Naturally, this spin network likely results in strong frustration.  
The frustration may be consistent with the strong reduction of the effective exchange 
interaction and ordering  
temperature $T_N$. 
Note that in the case of competing ferromagnetic and antiferromagnetic interactions the ratio 
$T_N/\Theta$ does not provide reliable information on magnetic frustration. 
Since Co$^{2+}$ ions behave more like paramagnets in \tetra, 
one may argue that the frustration effect modifies the 
spin dynamics which causes the peculiar upturn of \slr. 
At low temperatures, the frustration can induce the incommensurate or 
spin-glass-like magnetic ordering. 
In this case, magnetic short-range fluctuations may extend far above $T_N$, 
being responsible for the additional enhancement of \slr\ which was observed 
below $\sim150$ K. 

Now we discuss the puzzling feature of the $1/T$-dependence of \slr, 
which implies the inverse quadratic temperature dependence of the dynamical  
susceptibility. To begin with, one may conjecture that Li diffusion motion causes the 
increase of \slr\ with decreasing $T$.  However, we rule out this possibility because both 
mobile \li\ and immobile \pho\ nuclei detect the identical $T$-dependence of 
\slr, as demonstrated in Fig. \ref{fig:T1comp}. In principle, 
$A_\text{hf}$ may increase with decreasing $T$, causing the $T$-dependence of 
\slr. Again, this is clearly not the case from the uniquely defined $A_\text{hf}$ from the 
$\mathcal{K}$ vs. $\chi$ plot over a wide temperature range (see the inset of 
Fig. \ref{fig:31K}). 
Therefore, we conclude that unusual spin dynamics is 
present and persists even  
in the high $T$ region ($T\gg 10J_\text{ex}$), causing the inverse quadratic 
$T$-dependence of $\sum_q\chi''(q,\omega_L)$. 
One plausible explanation could be given if $J_\text{ex}$ in Eq. (\ref{eq:2}) 
decreases with decreasing temperature.\cite{baek10b} Although this scenario may 
be incompatible with the well-defined $\Theta\propto J_\text{ex}$, if spin 
fluctuations at small $q<Q$ are developed with decreasing temperature,
the resultant effective exchange coupling could be reduced.

\section{Conclusions}

We present \li\ and \pho\ NMR studies in both non-olivine and olivine 
structured LiCoPO$_4$ microcrystals. 
It turns out that the exchange  
interactions among the Co$^{2+}$ spins are greatly reduced in \tetra,
which accounts for the difference of the spin-lattice relaxation rates \slr\ between 
the two compounds.
In contrast to the Curie-Weiss behavior of the static susceptibility found at high 
temperatures,  the  
dynamical spin susceptibility deduced from the spin-lattice  
relaxation rates is inversely quadratic in temperature, which is particularly 
strong and robust in \olivine. For \tetra, the unusual temperature dependence 
is considerably weakened and breaks down at low temperatures.   
Together with the reduced effective exchange coupling and ordering 
temperature, this different spin dynamics is attributed to strong 
frustration effect inherent in the  
corner-shared CoO$_4$-PO$_4$ geometry of this metastable material.  
The additional enhancement \slr\ at low temperatures in \tetra\ suggests that the 
frustration may lead to complex incommensurate magnetic order.  

\section*{Acknowledgement}
We thank Andrei Malyuk for annealing the \tetra\ sample. 
This work was supported by the DFG (Grants
No. GR3330/3-1 and No. KL1824/2-2) and by the BMBF (Project No. 03SF0397). S.B. 
acknowledges support by DFG Research Grant No. BA 4927/1-1.
\bibliography{mybib}

\end{document}